\documentclass[twocolumn,showpacs,amsmath,amssymb,prd]{revtex4}

\usepackage{graphicx}
\usepackage{dcolumn}
\usepackage{bm}
\usepackage{epsf}

\begin{document}

\title{Illustrating Stability Properties of Numerical
	Relativity in Electrodynamics}

\author{A. M. Knapp${}^1$, E. J. Walker${}^1$ and T. W. Baumgarte${}^{1,2}$}

\affiliation{${}^1$ Department of Physics and Astronomy, Bowdoin College,
	Brunswick, ME 04011}

\affiliation{${}^2$ Department of Physics, University of Illinois at
	Urbana-Champaign, Urbana, IL, 61801}

\begin{abstract}
We show that a reformulation of the ADM equations in general
relativity, which has dramatically improved the stability properties
of numerical implementations, has a direct analogue in classical
electrodynamics.  We numerically integrate both the original and the
revised versions of Maxwell's equations, and show that their distinct
numerical behavior reflects the properties found in linearized general
relativity.  Our results shed further light on the stability
properties of general relativity, illustrate them in a very
transparent context, and may provide a useful framework for further
improvement of numerical schemes.
\end{abstract}


\maketitle

Motivated by the prospect of gravitational wave detections and the
accompanying need for theoretical gravitational wave templates, much
effort has recently gone into the development of numerical relativity
algorithms that are capable of modeling the most promising sources of
gravitational radiation, in particular the inspiral and coalescence of
binary black holes and neutron stars.  In the past, progress has been
hampered by numerical instabilities that arise in straight-forward
implementations of the traditional Arnowitt-Deser-Misner (ADM,
\cite{adm62}) $3+1$ decomposition of Einstein's equations
(e.g.~\cite{bmss95,aetal98}).  These instabilities have been associated
with the mathematical structure of the ADM equations, and as a cure a
number of hyperbolic formulations have been suggested
(e.g.~\cite{bmss95,ay99} as well as \cite{r98} and references therein).
Alternatively, Shibata and Nakamura \cite{sn95} and later Baumgarte
and Shapiro \cite{bs99} suggested a reformulation of the ADM equations
that has been demonstrated to dramatically improve the stability
properties of numerical implementations (e.g.~\cite{comparisons}).
While the exact reason for this improvement is still somewhat
mysterious (see \cite{aabss00}, hereafter AABSS, and \cite{math}),
this new formulation, now often called the BSSN formulation, is quite
widely used.

In this Brief Report we show that the BSSN reformulation of the ADM
equations has a direct analogue in classical electrodynamics (E\&M).
We numerically implement both the original and the revised versions of
Maxwell's equations, and find that their distinct numerical properties
reflect those found in general relativity (GR).  As suggested by
AABSS, these properties can be identified with the propagation of
constraint violating modes.  We present our findings in the hope that,
in addition to being a very transparent illustration of the stability
properties of GR, they may prove useful for the future development and
improvement of numerical algorithms.

In a $3+1$ decomposition of GR, Einstein's equations split into the
two constraint equations
\begin{equation} \label{ham1}
R - K_{ij}K^{ij} + K^2 = 2 \rho
\end{equation}
and
\begin{equation} \label{mom1}
D_j K^{j}_{~i} - D_i K = S_i
\end{equation}
and the two evolution equations
\begin{equation} \label{gdot1}
d_t \gamma_{ij} = - 2 \alpha K_{ij}
\end{equation}
and
\begin{equation} \label{Kdot1}
d_t K_{ij} = - D_i D_j \alpha + \alpha ( R_{ij} 
        - 2 K_{il} K^l_{~j} + K K_{ij} - M_{ij} ). 
\end{equation}
Here $\gamma_{ij}$ is the spatial metric, $K_{ij}$ the extrinsic
curvature, $K = \gamma^{ij} K_{ij}$ its trace, $\alpha$ and
$\beta^i$ are the lapse function and the shift vector, and $\rho$, $S_i$
and $M_{ij}$ are matter source terms.  The time derivative is defined
as $d_t = \partial_t - {\mathcal L}_{\beta}$, and $R_{ij}$
\begin{eqnarray} \label{ricci}
R_{ij} & = & - \frac{1}{2} \gamma^{kl} 
	\Big( \gamma_{ij,kl} + \gamma_{kl,ij}
	- \gamma_{kj,il} - \gamma_{il,kj} \Big)\\
 	& & + \gamma^{kl} \Big( \Gamma^m_{il} \Gamma_{mkj}
 	- \Gamma^m_{ij} \Gamma_{mkl} \Big), \nonumber
\end{eqnarray}
$D_i$ and $\Gamma^i_{jk}$ are the Ricci tensor, the covariant
derivative, and the connection coefficients associated with
$\gamma_{ij}$.  Finally, $R$ is the scalar curvature $R = \gamma^{ij}
R_{ij}$.  Equations (\ref{ham1}) through (\ref{Kdot1}) are commonly
refered to as the ADM equations \cite{adm62}.

The first term in the Ricci tensor (\ref{ricci}) is an
elliptic operator acting on the components of the spatial metric
$\gamma_{ij}$.  If the Ricci tensor contained only this term, the two
evolution equations (\ref{gdot1}) and (\ref{Kdot1}) could be combined
to form a wave equation for $\gamma_{ij}$.  This property is spoiled
by the appearance of the three other second derivative terms in
(\ref{ricci}), suggesting that these terms may be responsible for the
appearance of instabilities in many straight-forward,
three-dimensional implementations of the ADM equations.  This problem
can be avoided by either using a hyperbolic formulation of GR
(e.g.~\cite{r98}), or by eliminating the mixed second derivatives as
in the BSSN formulation.

In this formulation, the conformally related metric $\bar \gamma_{ij}$
is defined as $ \bar \gamma_{ij} = e^{- 4 \phi} \gamma_{ij}, $ where
the conformal factor $e^{\phi}$ is chosen so that the determinant of
$\bar \gamma_{ij}$ is unity, $\bar \gamma = 1$.  The conformal
exponent $\phi$ as well as the trace of the extrinsic curvature, $K$,
are evolved as independent variables.  Their evolution equations can
be found from the traces of equations (\ref{gdot1}) and (\ref{Kdot1}),
while the trace-free parts of those equations form evolution equations
for $\bar \gamma_{ij}$ and the trace-free part of the extrinsic
curvature, $\bar A_{ij}$.  The latter equation still contains the
Ricci tensor $\bar R_{ij}$ associated with $\bar \gamma_{ij}$, which
contains all the mixed second derivatives of (\ref{ricci}).  The
crucial step is to realize that these second derivatives can be
absorbed in a first derivative of the ``conformal connection
functions''
\begin{equation} \label{ccf}
\bar \Gamma^i \equiv \bar \gamma^{jk} \bar \Gamma^{i}_{jk}
	= - \bar \gamma^{ij}_{~~,j},
\end{equation}
where the last equality holds because $\bar \gamma = 1$.  Here and in
the following all barred quantities are associated with $\bar
\gamma_{ij}$.  In terms of $\bar \Gamma^i$, the Ricci tensor can be
written
\begin{eqnarray} \label{ricci2}
\bar R_{ij} & = & -  \frac{1}{2} \bar \gamma^{lm}
	\bar \gamma_{ij,lm} 
	+ \bar \gamma_{k(i} \partial_{j)} \bar \Gamma^k
	+ \bar \Gamma^k \bar \Gamma_{(ij)k}  + \\
	& & \bar \gamma^{lm} \left( 2 \bar \Gamma^k_{l(i} 
	\bar \Gamma_{j)km} + \bar \Gamma^k_{im} \bar \Gamma_{klj} 
	\right) \nonumber
\end{eqnarray}
(compare \cite{old_guys}).  Evidently, the only remaining second
derivative term is the elliptic operator $\bar \gamma^{lm} \bar
\gamma_{ij,lm}$, if the $\bar \Gamma^i$ are considered as independent
functions.  For that purpose, an evolution equation is derived by
permuting a time and space derivative in (\ref{ccf})
\begin{equation} 
\partial_t \bar \Gamma^i
=  - \partial_j \Big( 2 \alpha \bar A^{ij} 
	- 2 \bar \gamma^{m(j} \beta^{i)}_{~,m}
	+ \frac{2}{3} \bar \gamma^{ij} \beta^l_{~,l} 
	+ \beta^l \bar \gamma^{ij}_{~~,l} \Big).
\end{equation}
The divergence of the extrinsic curvature can now be eliminated with 
the help of the momentum constraint (\ref{mom1}), which yields the
evolution equation
\begin{eqnarray} \label{Gammadot}
&& \partial_t \bar \Gamma^i =
	- 2 \bar A^{ij} \partial_j \alpha \\
&& ~~	+ 2 \alpha \Big(
	\bar \Gamma^i_{jk} \bar A^{kj} 
	- \frac{2}{3} \bar \gamma^{ij} \partial_j K
	- \bar \gamma^{ij} S_j + 6 \bar A^{ij} \partial_j \phi \Big) 
	\nonumber \\
&& ~~ + \beta^j \bar \partial_j \Gamma^i 
	- \bar \Gamma^j \partial_j \beta^i
	+ \frac{2}{3} \bar \Gamma^i \partial_j \beta^j 
	+ \frac{1}{3} \bar \gamma^{li} \beta^j_{,jl} 
	+ \bar \gamma^{lj} \beta^i_{,lj}. \nonumber
\end{eqnarray}
As in \cite{bs99} we will refer to the ADM equations (\ref{ham1})
through (\ref{Kdot1}) as System I, and to the new BSSN system of
equations as System II.  A complete listing and derivation of the
latter can be found in \cite{bs99}.

\begin{figure*}[t]
\begin{center}
\leavevmode
\epsfxsize=2.2in
\epsffile{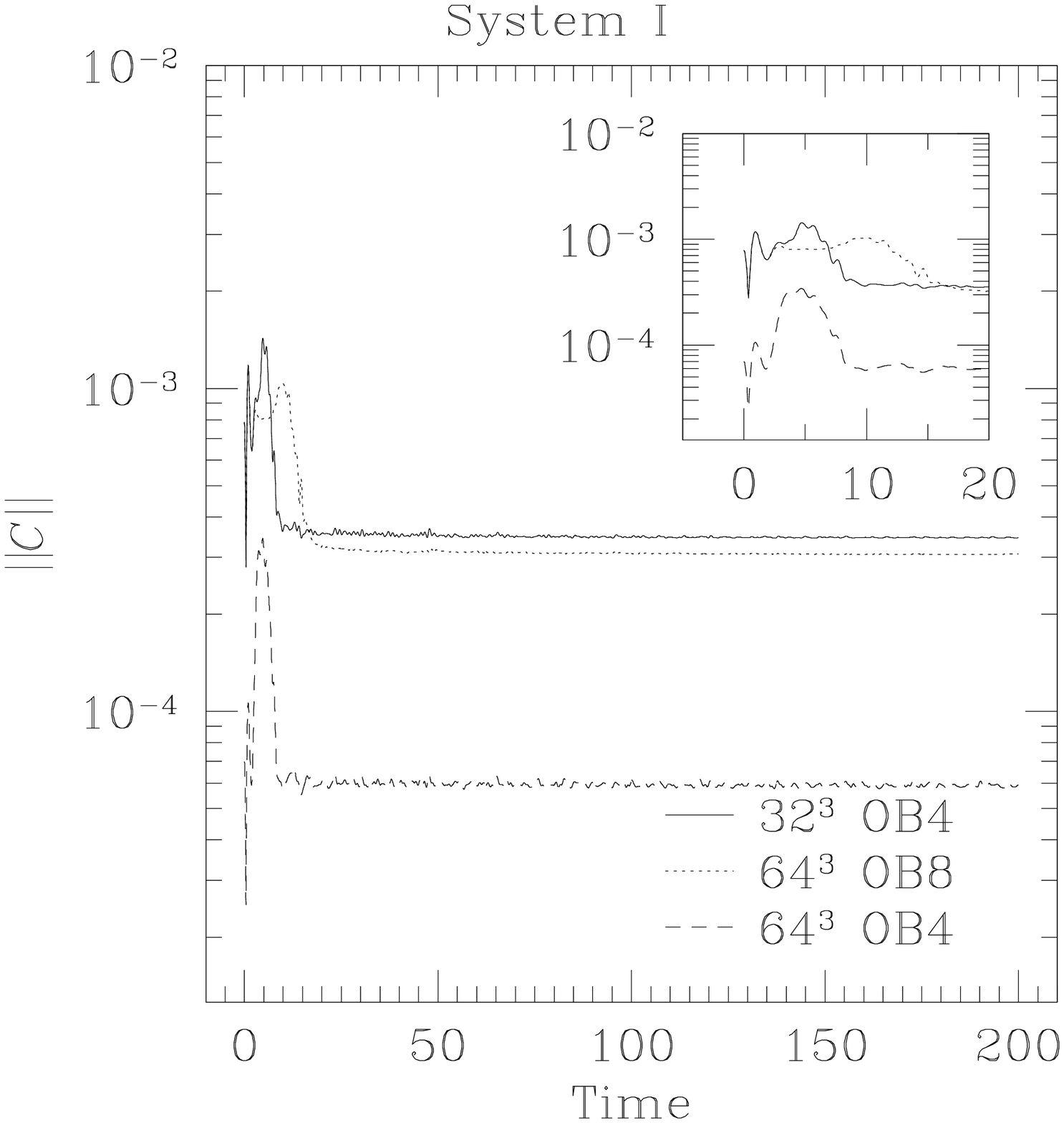}~~~~~~~~
\epsfxsize=2.2in
\epsffile{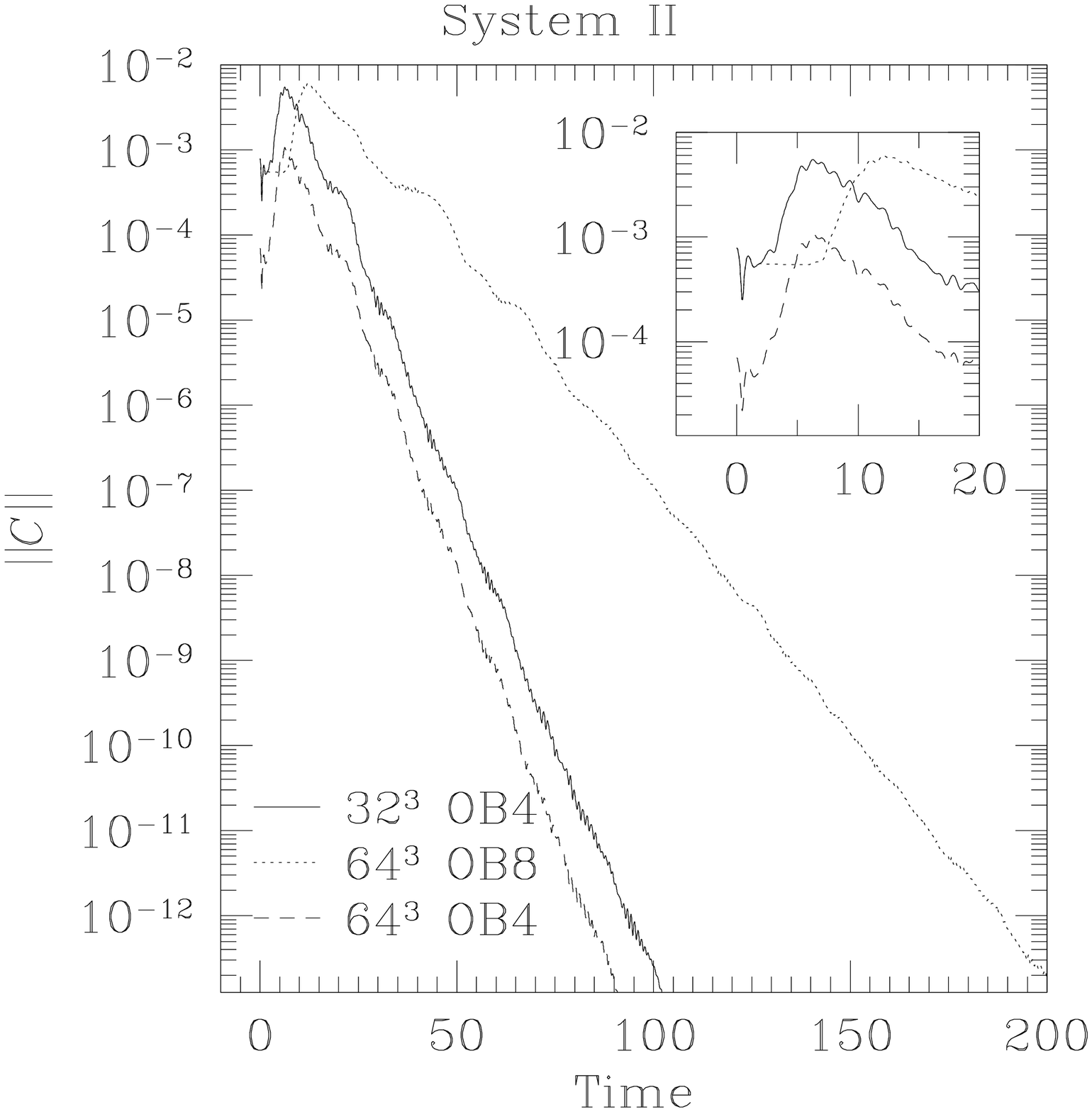}
\end{center}
\caption{The integrated constraint violation $\|\mathcal{C}\| \equiv
(\int \mathcal{C}^2 dV)^{1/2}$ for evolutions using Lorentz gauge with
resolution $32^3$ and $64^3$ with outer boundaries at 4 (OB4) and 8
(OB8).}
\label{fig1}
\end{figure*}

In an effort to better understand the improved numerical behavior of
System II, AABSS linearized Systems I and II and identified their
characteristic structure.  They found that System I has constraint
violating modes with a characteristic speed of zero.  In System II,
the characteristic speed of these modes changes to the speed of light.
AABSS further demonstrated that in a non-linear model problem the
existence of non-propagating, constraint violating modes may lead to
numerical instabilities, as encountered in implementations of System
I.  Their analysis also demonstrated that the usage of the momentum
constraint in the derivation of (\ref{Gammadot}) is crucial for the
characteristic speed for the constraint-violating mode to change to a
non-zero value, and hence for the stability of the system.
In the following we will show that very similar properties can be
found in E\&M.

In terms of a vector potential $A_i$, Maxwell's equations can be 
written as the evolution equations
\begin{eqnarray}
\partial_t A_i & = & - E_i - D_i \psi \label{Adot1} \\
\partial_t E_i & = & - D^j D_j A_i + D^j D_i A_j - 4 \pi j_i, \label{Edot1}
\end{eqnarray}
and the constraint equation
\begin{equation}
D_i E^i = 4 \pi \rho_e. \label{divE}
\end{equation}
Here $E^i$ is the electrical field, $\rho_e$ the charge density, $j_i$
the flux, and $\psi$ the scalar gauge potential.  Identifying $A_i$ with
$\gamma_{ij}$, $E_i$ with $K_{ij}$, and $\psi$ with $\beta^i$, we see
that the structure of the evolution equations (\ref{Adot1}) and
(\ref{Edot1}) is very similar to that of equations (\ref{gdot1}) and
(\ref{Kdot1}), and that the constraint equation (\ref{divE}) can 
be similarly identified with the momentum constraint (\ref{mom1}).
In analogy with GR, we refer to equations (\ref{Adot1}) through
(\ref{divE}) as System I.

In further analogy, we will eliminate the mixed second derivative in
(\ref{Edot1}) by introducing a new variable $\Gamma$
\begin{equation}
\Gamma \equiv D_i A^i \label{Gamma}
\end{equation}
(compare (\ref{ccf})), in terms of which (\ref{Edot1}) becomes
\begin{equation}
\partial_t E_i = - D^j D_j A_i + D_i \Gamma - 4 \pi j_i.  \label{Edot2}
\end{equation}
As in (\ref{ricci2}), the mixed second derivatives have been absorbed
in a first derivative of the new variable.  An evolution equation for
$\Gamma$ can again be derived by permuting a time and space derivative
in the definition of the new variable (\ref{Gamma})
\begin{equation}
\partial_t \Gamma
	= D_i \partial_t A^i = - D_i E^i - D^i D_i \psi = - 4 \pi\rho_e  
	- D^i D_i \psi. \label{Gammadot2}
\end{equation}
Here we have used the constraint (\ref{divE}),
which, as in GR, will turn out to be crucial (see (\ref{Cwave})
below).  Equations (\ref{Adot1}), (\ref{Edot2}), (\ref{Gammadot2})
form the evolution equations of what we call System II.  The
definition (\ref{Gamma}) together with (\ref{divE}) form the
constraint equations of System II.

For vanishing scalar potential $\psi = 0$, an analytical vaccuum
solution to Maxwell's equations is given by the purely toriodal dipole
field
\begin{equation}
A^{\hat{\phi}} = {\mathcal A} \sin \theta \left( \frac{e^{-\lambda
v^{2}} - e^{-\lambda u^{2}}}{r^2} - 2 \lambda \frac{v e^{-\lambda
v^{2}} + u e^{- \lambda u^{2}}}{r}
\right). \label{Aanal}
\end{equation} 
Here ${\mathcal A}$ is the amplitude, $\lambda$ parameterizes the size
of the wavepackage, and $u$ and $v$ are the retarded and advanced time
$u \equiv t+r$ and $v \equiv t-r$.  According to (\ref{Adot1}),
$E^{\hat \phi}$ can be found by taking a time derivative of
(\ref{Aanal}).  Since $\rho_e = 0$ and $D_i A^i = 0$ in this solution,
$\psi = 0$ is consistent with both the Coulomb gauge and the Lorentz
gauge
\begin{equation} \label{psidot}
\partial_t \psi = - D_i A^i - 4\pi\rho_e = - \Gamma - 4\pi\rho_e
\end{equation}
(where the second equality applies for System II), if appropriate
boundary and initial conditions are chosen.  All results shown below
were obtained with Lorentz gauge.

As initial data for our dynamical simulations we adopt the analytical
solution (\ref{Aanal}) at $t=0$
\begin{equation}
A^{\hat{\phi}} = 0, ~~~ 
E^{\hat{\phi}} = 8 {\mathcal A} r \sin \theta \lambda^2 e^{-\lambda r^2}
\end{equation}
with $\mathcal{A} = \lambda = 1$ and transformed into cartesian
coordinates.

We numerically implement System I and II following the algorithm of
\cite{bs99} as closely as possible.  In particular, we wrote the code
in three spatial dimensions using cartesian coordinates.  We use an
iterative Crank-Nicholson scheme \cite{t00} to update the evolution
equations, and impose outgoing wave boundary conditions on $E^i$,
$A^i$ and $\Gamma$ as in \cite{bs99}.  We verified that the numerical
solution converges to the analytical solution to second order as long
as the solution is not affected by the outer boundaries (which are
first order accurate).

We compare the performance of System I and II by monitoring the 
constraint violation
\begin{equation}\label{C}
\mathcal{C} \equiv D_i E^i - 4\pi\rho_e.
\end{equation}
In Fig.~\ref{fig1}, we show integrated values
$\|\mathcal{C}\| \equiv (\int \mathcal{C}^2 dV)^{1/2}$ for Systems I
and II for two different gridsizes ($32^3$ and $64^3$) and two
different locations of the outer boundary (at 4 (OB4) and 8 (OB8)).

\begin{figure*}[t]
\begin{center}
\leavevmode
\epsfxsize=2.3in
\epsffile{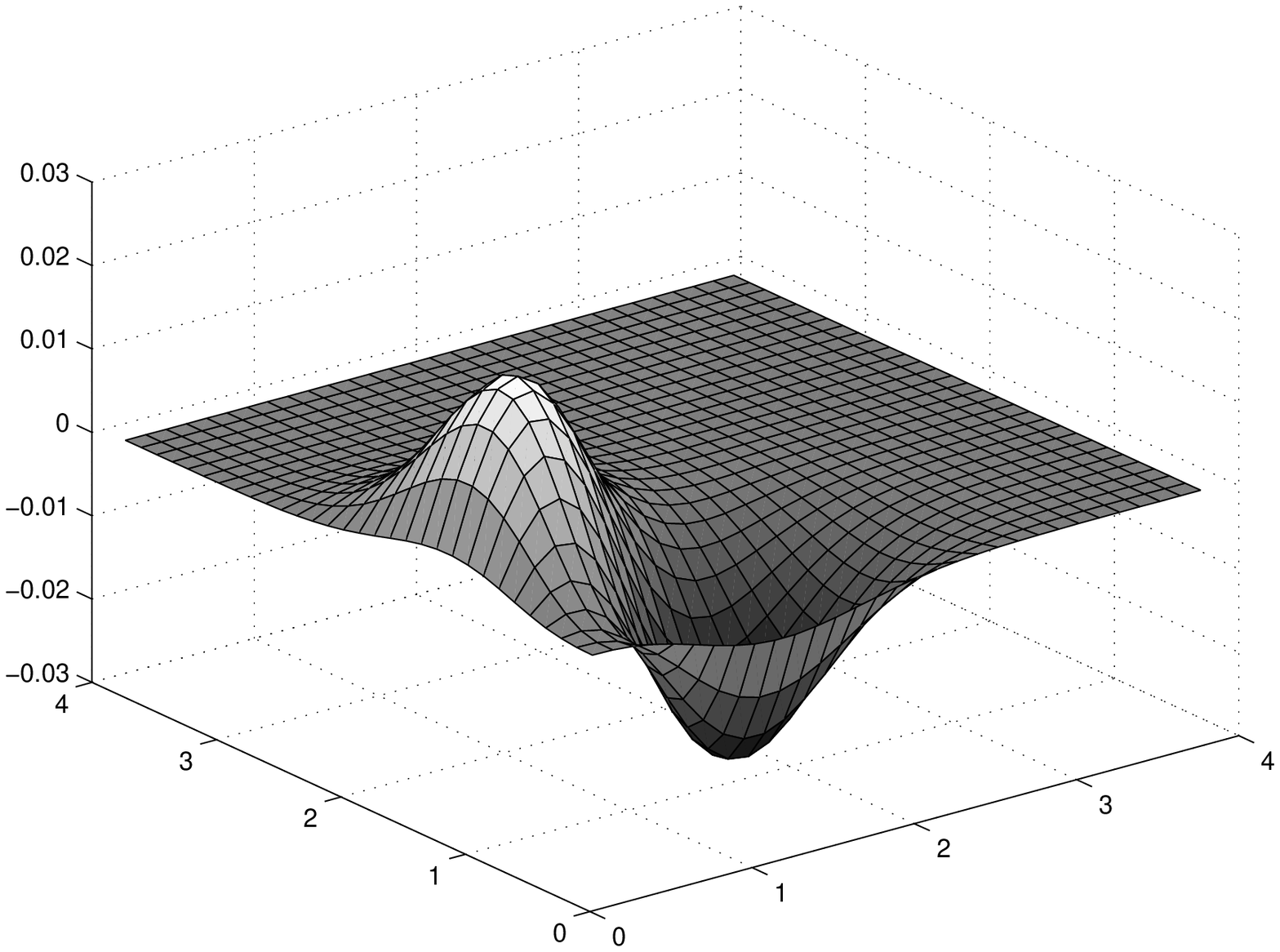}~~~~~~~~
\epsfxsize=2.3in
\epsffile{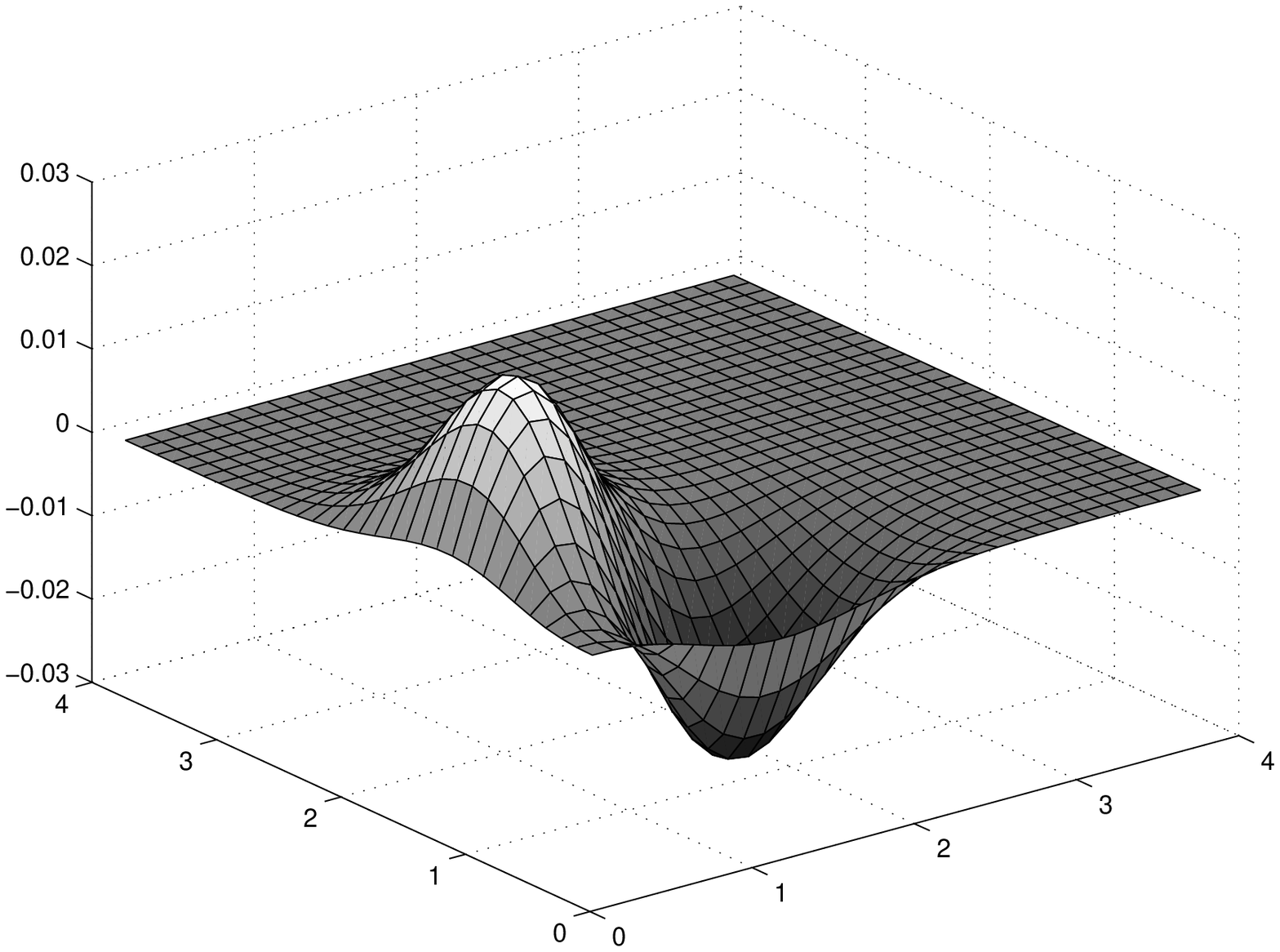} \\
\leavevmode
\epsfxsize=2.3in
\epsffile{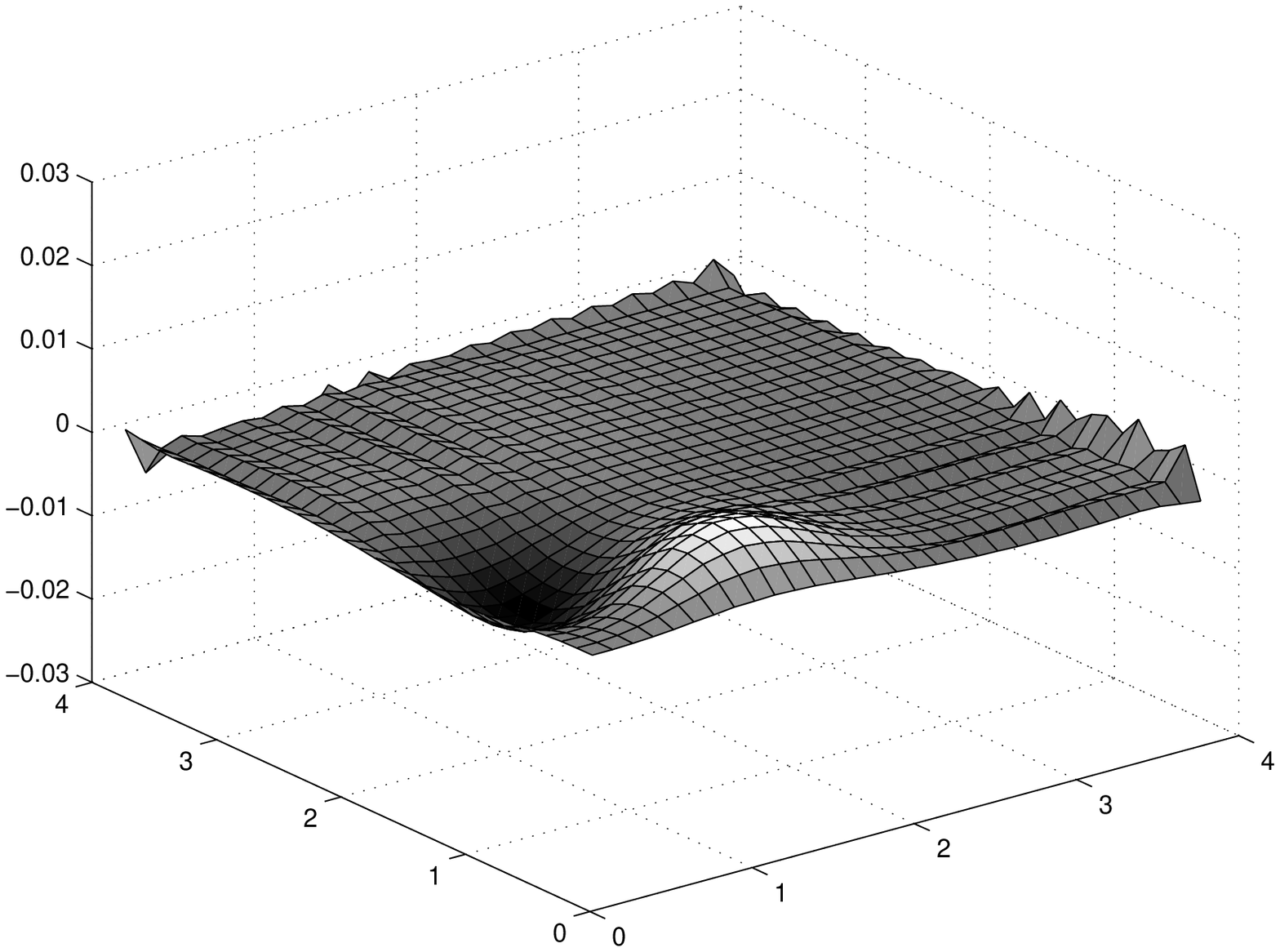}~~~~~~~~
\epsfxsize=2.3in
\epsffile{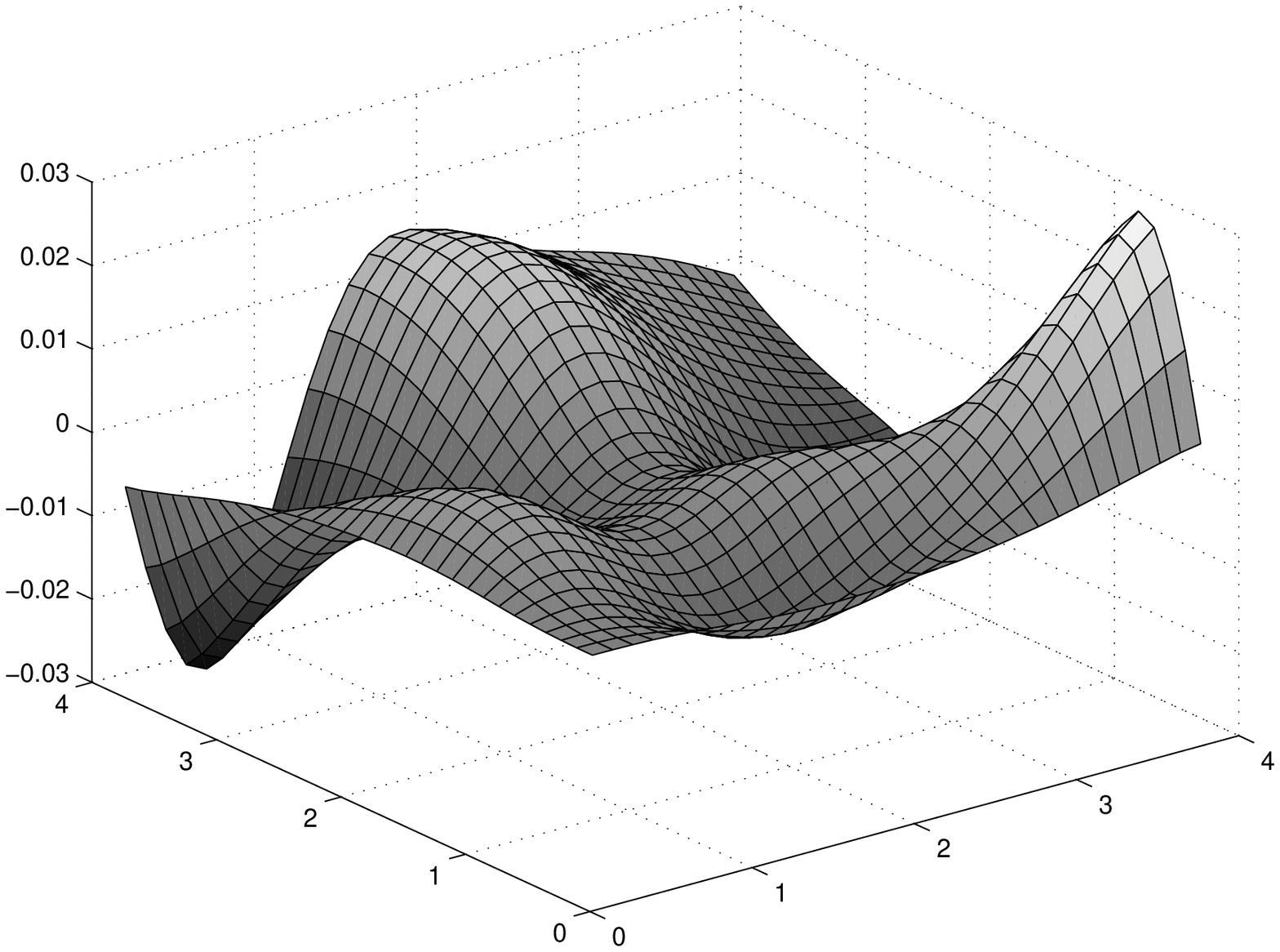}\\
\leavevmode
\epsfxsize=2.3in
\epsffile{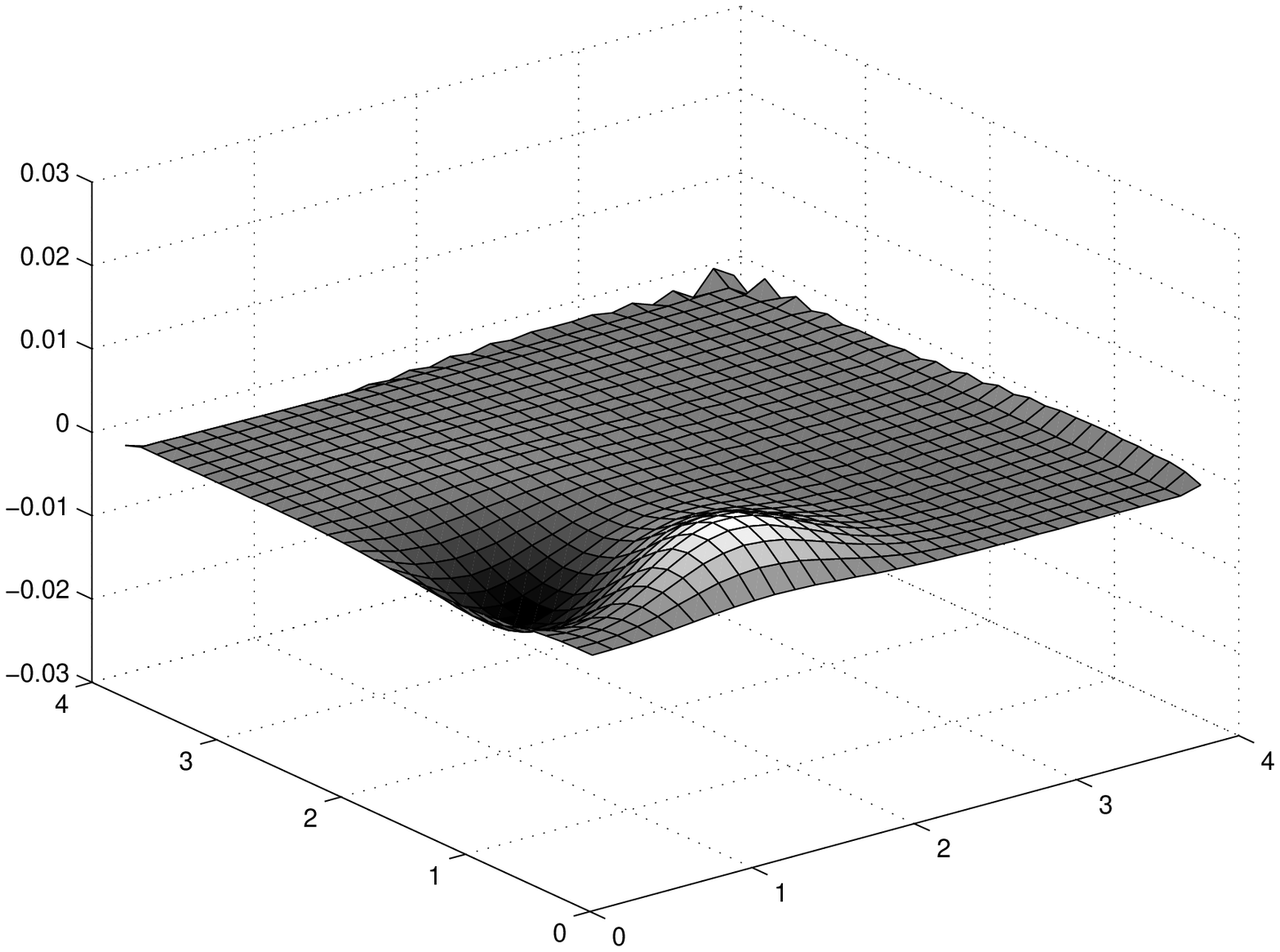}~~~~~~~~
\epsfxsize=2.3in
\epsffile{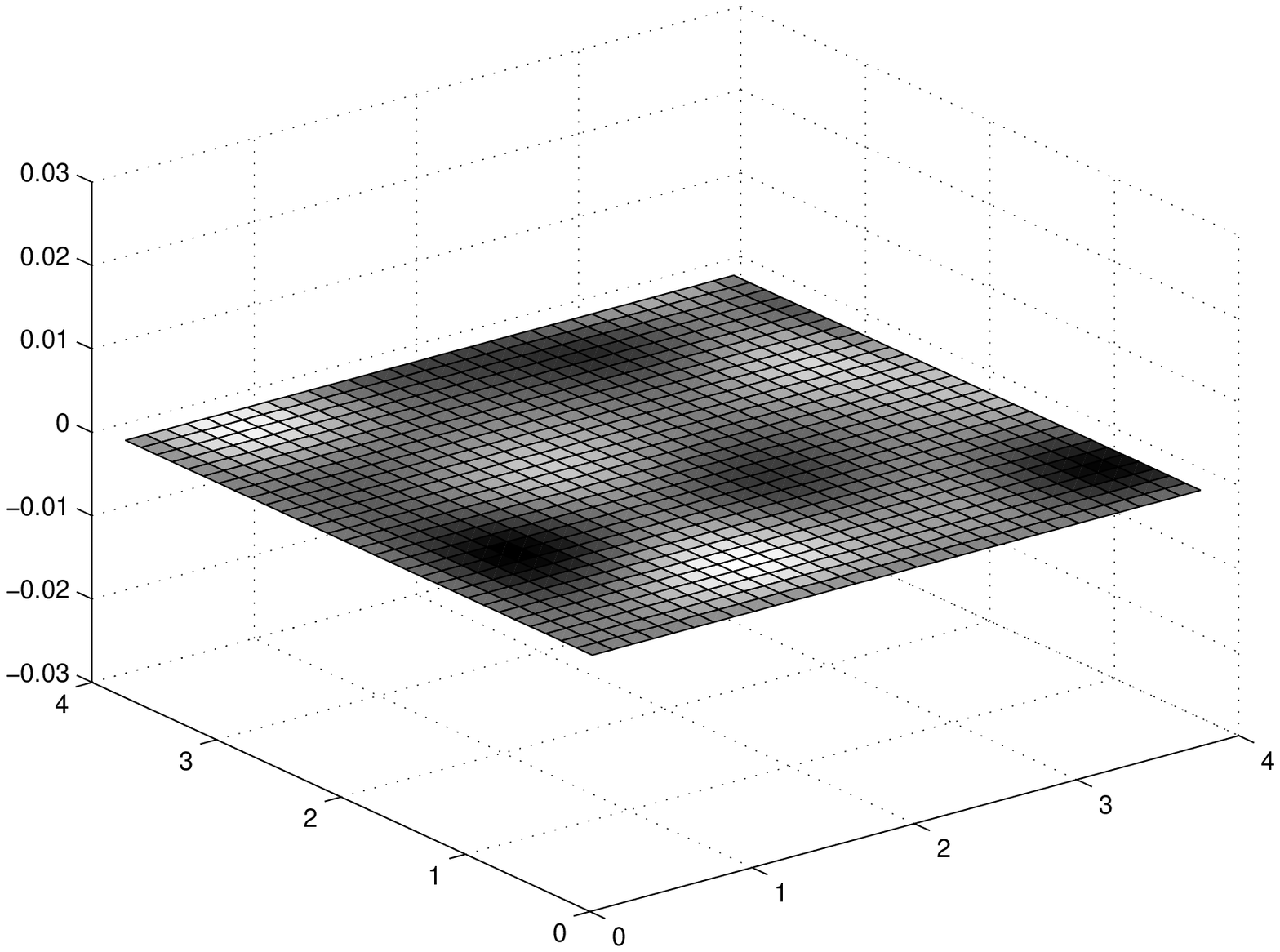}
\end{center}
\caption{${\mathcal C}$ in the z = .0625 plane (the first of our grid
points) for System I (left column) and System II (right column) for a
$32^3$ OB4 evolution in Lorentz gauge.  We show results for $t=0$ (top
row), $t=9.375$ (middle row) and $t=100$ (bottom row).  ${\mathcal C}$
settles down to a constant profile in System I, but dissipates away in
System II.}
\label{fig2}
\end{figure*}

At early times, System I violates the constraint (\ref{divE}) to a
lesser degree than System II.  After about a light-crossing time, when
the electro-magnetic wave has left the numerical grid,
$\|\mathcal{C}\|$ settles down to a nearly constant value, which
primarily depends on the grid-resolution.  In System II,
$\|\mathcal{C}\|$ is also largest after about a light-crossing time,
but after that $\|\mathcal{C}\|$ decreases exponentially.  As one
might expect, the decay time of this exponential fall-off scales with
the location of the outer boundary.

To further compare these Systems, we compare snapshots of the
constraint violation ${\mathcal C}$ for the $32^3$ OB4 evolution in
Fig.~\ref{fig2}.  With identical initial data, both Systems have
identical values for ${\mathcal C}$ at $t=0$.  At an intermediate time
${t=9.375}$, $\mathcal{C}$ is larger in System II than in System I, as
one expects from Fig.~\ref{fig1}.  At a much later time (${t=200}$),
however, System I has settled down to a constant shape, while
$\mathcal{C}$ in System II has almost completely dissipated.

These numerical results demonstrate that, as in GR, the constraint
violation ${\mathcal C}$ behaves very differently in the two systems.
In E\&M, this different behavior can be understood very easily from an
analytic argument.

For System I, it is easy to show that a time derivative of the 
constraint violation (\ref{C}) vanishes
\begin{equation} 
\partial_t \mathcal{C} = D_i \partial_t E^i - 4 \pi \partial_t \rho_e
	= - 4 \pi (D_i j^i + \partial_t \rho_e) = 0,
\end{equation}
where we have used the continuity equation $D_i j^i + \partial_t
\rho_e = 0$.  This explains why at late times the profile of
${\mathcal C}$ remains unchanged in System I.  This property is the
analogue of AABSS's finding that the linearized ADM equations have
non-propagating, constraint violating modes.

For System II, on the other hand, it can be shown that 
${\bf \mathcal C}$ satisfies a wave equation
\begin{eqnarray} \label{Cwave}
\partial^2_t {\mathcal C} & = & 
        \partial_t D^i \partial_t E_i - 4 \pi \partial^2_t \rho_e \nonumber \\
        & = & \partial_t D^i (- D_j D^j A_i + D_i \Gamma  - 4 \pi j_i)
         - 4 \pi \partial^2_t \rho_e \nonumber \\
        & = &  - D^i ( D_j D^j \partial_t A_i  
        - D_i \partial_t \Gamma ) 
        - 4 \pi \partial_t (D^i j_i + \partial_t \rho_e) \nonumber  \\
        & = &  D^i \Big( D_j D^j (E_i + D_i \psi) 
        - D_i (4 \pi \rho_e + D^j D_j \psi) \Big)  \nonumber  \\
        & = &  D_j D^j (D^i E_i - 4 \pi \rho_e) =  D_j D^j {\mathcal C},
\end{eqnarray}
which explains why the constraint violations propagate away in System
II.  This is the analogue of AABSS's finding that in the relativistic
System II (the BSSN equations) the constraint violating modes
propagate with the speed of light.  Moreover, we now realize why the
usage of the constraint (\ref{divE}) in (\ref{Gammadot}) was crucial
-- without having made this substitution the terms on the last line of
(\ref{Cwave}) would have cancelled, $\partial_t^2 {\mathcal C} = 0$,
leading to a non-propagating constraint violation as in System I.
The addition of constraint equations to the evolution equations has
been found to be crucial in several other formulations of Einstein's
equations as well (e.g.~\cite{ay99,constraint}).

To summarize, we see that the numerical stability properties of two
formulations of GR are beautifully reflected in similar formulations of
E\&M.  Maxwell's equations therefore provide a very transparent
framework for analyzing these properties, which may be useful for
future algorithm development.  For example, Fig. \ref{fig2} shows that
the outer boundaries produce much more noise in System I than in
System II, which points to an inconsistency between the treatment of
the interior equations and the boundary.  Similar problems are likely
to occur in GR, but might be easier to analyze in the simpler
framework of E\&M.

\acknowledgments
 
This paper was supported in part by NSF Grant PHY 99-02833 to the
University of Illinois at Urbana-Champaign and Bowdoin College as a
subrecipient.  AMK gratefully acknowledges support from the Mellon
Foundation.

\end{document}